# 2π-Steradian, Energetic-Ion Sensor

Donald G. Mitchell
The Johns Hopkins University Applied Physics Laboratory
11100 Johns Hopkins Road, Laurel, MD  20723-6099

# INTRODUCTION

Because energetic particles populate both planetary magnetospheres and interplanetary space in significant quantities, energetic-ion sensors have been flown since the beginning of the space age. Early sensors were solid-state detector (SSD) telescopes, with conical fields of view, often swept through a circle by virtue of the spin motion of the spacecraft (eg, IMP 7 and 8, ISEE 1 and 2, etc.). In the 1980s and 1990s, foil/microchannel plate (MCP) time-of-flight (TOF) measurements were added to the energy measurement provided by the SSD (eg, AMPTE/CCE MEPA, Geotail EPIC/ICS, Galileo EPD). The resulting energy and velocity uniquely identified ion mass. More recently, we have developed a 2-D fan acceptance angle sensor that includes both energy and TOF. When mounted on a spinning spacecraft, this 160° x 12° FOV sweeps out nearly $4\pi$ steradians in one spin. This sensor, dubbed the "hockey puck" for its shape, is currently in flight on MESSENGER (EPS) and New Horizons Pluto (PEPPSI).

Increasingly, and particularly in the planetary program where low launch frequency demands multiple science-goal payloads, energetic-ion sensors fly on 3-axis stabilized spacecraft (e.g., MESSENGER EPS, New Horizons (Pluto) PEPPSI, Cassini MIMI (Livi et al. 2003; Krimigis et al., 2004). While 3-axis stabilization serves imaging science well, it hampers the goal of obtaining $4\pi$-steradian ion measurements. In some cases, articulation mechanisms have been added to provide an additional degree of freedom, but this is expensive in mass and power, and increases risk (e.g., the Cassini MIMI/LEMMS turntable, which failed after one year in Saturn Tour).

Whereas the energetic-ion environment is rarely the sole motivation for a particular planetary mission, compelling science requirements have driven the inclusion of energetic-ion sensors on most past planetary missions (including all of those to magnetized planets), and will likely continue to do so in the future. The magnetospheres of the planets both trap and energize ions and electrons, whose sources are various: planetary atmospheres/ionospheres (Jupiter, Saturn, Uranus and Neptune), active moons (Io and possibly Europa at Jupiter, Enceladus at Saturn), moon atmospheres (Titan at Saturn, Triton at Neptune), sputtering of the surfaces of surfaces and rings (Mercury, Saturn, Jupiter, Neptune, Uranus), and the solar wind. These energetic particles further serve to modify the surfaces of moons and ring particles, influence the transport of material through charge exchange and sputtering, provide energy for atmospheric chemistry, and generate aurorae and synchrotron radiation.

The pitch angle distribution (PAD) of energetic particles is especially important in interpreting the measurements. Evolution of the PAD over time or location in a planetary magnetosphere is critical to unfolding particle sources and sinks, interactions with moons and rings, as well as acceleration and transport processes. Further complicating energetic-particle measurements, when measuring energetic-particle intensity from a 3-axis stabilized platform that makes attitude changes during a measurement (as, for example, when the spacecraft slews to train a telescope on a moon, or the planetary limb, or swings to point the high-gain antenna at Earth), a sensor with incomplete angular coverage will often see a large change in intensity. This change in intensity may be real, or it may be that the sensor look direction relative to the magnetic field changes during the maneuver, resulting in a change of the pitch angles of the particles accepted by the sensor. If the pitch angle distribution is not isotropic (which it rarely is), the change in intensity may simply reflect a gradient in the pitch angle distribution. Without full angular coverage, it may not be possible to unfold the change in intensity. Even if the change can be attributed to pitch angle structure, that structure will not be resolved, and the energetic-particle measurements will be incomplete and of reduced scientific value.

## A NEW APPROACH: THE "MUSHROOM"

We are developing an energetic-ion sensor that measures ion energy and composition, and covers $2\pi$ steradians without an articulation mechanism (the best that may reasonably be achieved from a body-mounted sensor). Two such heads, mounted on opposite sides of a spacecraft, can provide full $4\pi$-steradian coverage. The sensor will measure the angular distribution of ions with roughly 22.5-degree resolution, and with sufficient ion-mass resolution to characterize separately hydrogen, helium, CNO, neon, argon, and iron over an energy range of 30 keV to 2 MeV for protons and 70 keV to > 10 MeV for oxygen, as well as separating $He^3$ and $He^4$ over a reduced energy range (80 keV to 2 MeV). Separation of carbon, nitrogen, and oxygen from one another should be possible between about 0.3 and 10 MeV. The timing requirements are high, but no more stringent than others we have built to even as long ago as AMPTE. However, the design is mechanically somewhat complex, and requires development to a high precision breadboard level with ion beam test results so that a convincing flight proposal may be based on it. Based on its shape, we refer to this design as the "mushroom".

A cut through the proposed sensor is shown schematically in Figure 1. Details of some of the electron ray-tracing for the Start and Stop electrons are shown in Figures 2 and 3, respectively. Calculated species tracks are shown in Figure 4.

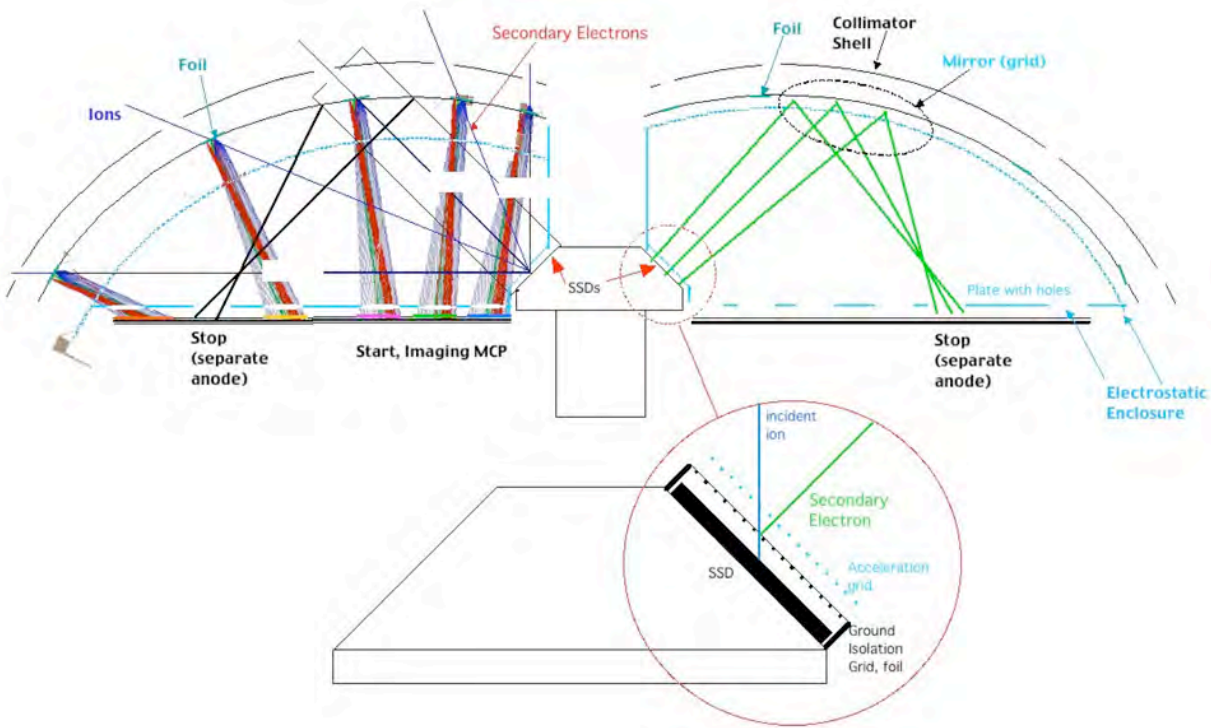

Figure 1. Mushroom schematic cut-away, illustrating ion and electron optics. The sensor is identical side to side, but different functionality is illustrated on each side. The left shows the basic concept for ion collimation (through holes in concentric shells) and secondary electron focusing (from entrance foils, across an equipotential volume,

onto a position-sensitive microchannel plate). The right shows the stop, secondary-electron optics. Both Start- and Stop-electron optics have been ray-traced in SIMION.

The SSDs lie on a central faceted cone. Stacked, perforated octagonal shell segments form the collimator, support thin foils, and comprise electrostatic, electron-optics grids (inner-most shell). Holes in the outer shell are aligned with small foils in the second shell to collimate ion trajectories on straight paths to the SSDs. The thin foils covering the holes in the middle shell eject secondary electrons as the ions penetrate them, and those secondary electrons are both accelerated and guided to a spot on the MCP by a bevel in the foil support shell and the grid surface of the innermost shell (see Figure 2). The secondary-electron position on the START MCP uniquely identifies the ion-entrance hole, and the pulse provides the start trigger for a TOF circuit. Secondary electrons generated as the ion strikes the surface of the SSD are accelerated through the equipotential volume and reflected into the STOP location on the MCP (Figure 3). This provides a stop impulse to the TOF circuit. The TOF is corrected for systematic time differences introduced by the geometry and the electron optics, and the energy, mass, and arrival direction are forwarded to the DPU where the ion mass is also calculated.

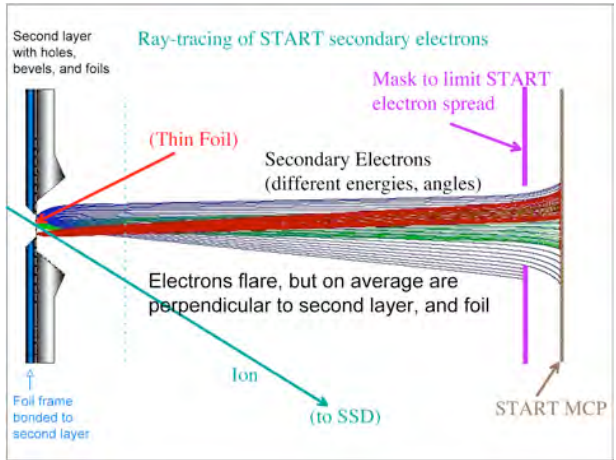

Figure 2. Ray-tracing results of lens assembly for Mushroom. Start, secondary electrons (<50 pico-sec dispersion). Electrons are mechanically blocked for perpendicular velocities above a prescribed maximum.

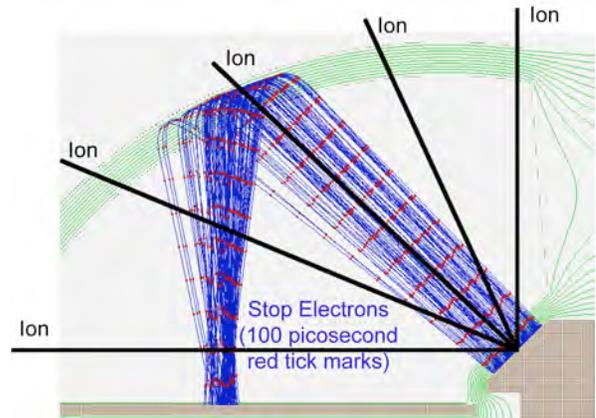

Figure 3. Ray-tracing results of lens assembly for Mushroom. Stop, secondary-electrons (<30 pico-sec dispersion). The Start and Stop signals are taken from separate anodes on the same MCPs.

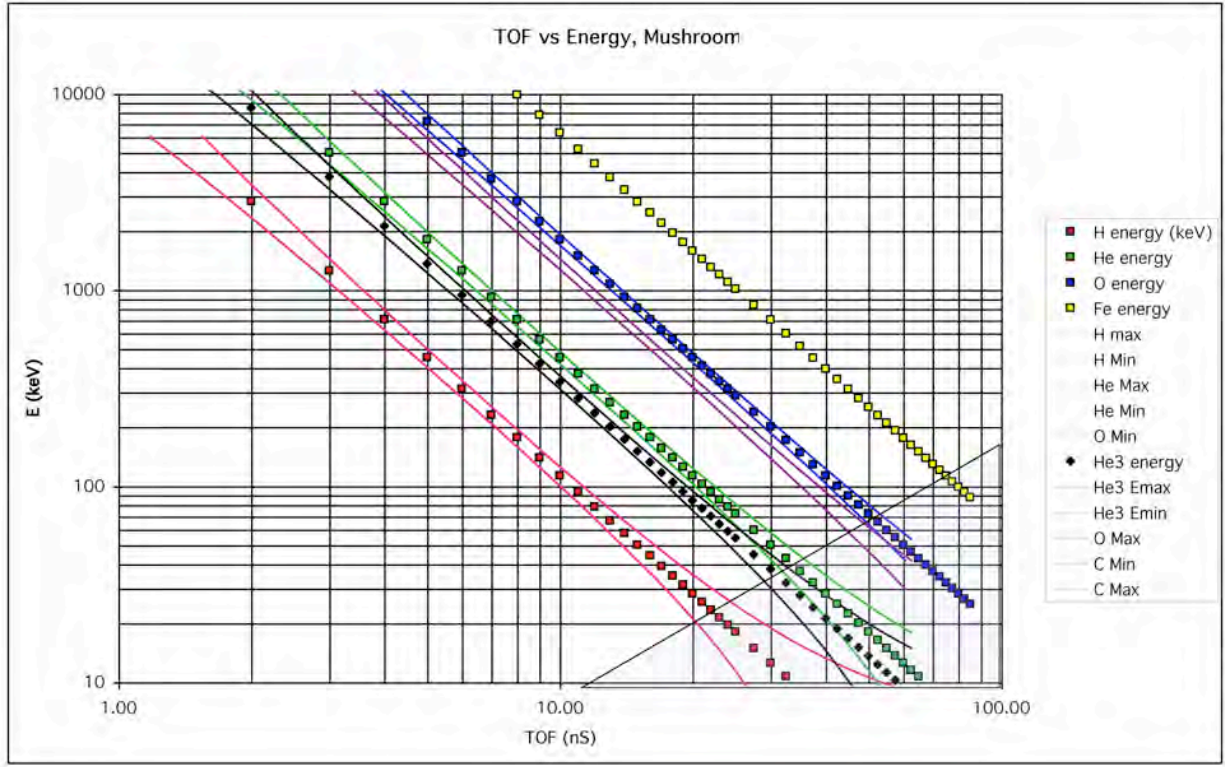

Figure 4. Mass separation for various ion species in Mushroom sensor design, including solid-state detector energy-resolution effects, secondary-electron time dispersion based on SIMION ray-tracing, ion path-length uncertainties, and A/D converter-bit resolution. Grey area (lower right) masks minimum SSD detection-energy limit.

Each octant is self-similar, with a separate SSD for ion energy and MCP for ion TOF measurement. The hole pattern alternates from one octant to the next, as shown in Figure 5. (The center-most holes, appearing only in every other octant, are redundant in that they all look parallel to the axis of symmetry.) The ions then pass through thin foils placed in a set of holes in a second shell, which completes their collimation. Secondary electrons generated on the foil exit surface are accelerated across a gap by a high potential on a grid that forms a third shell. The electrons coast through an equipotential volume, pass through a second grid and strike a MCP where their position and time is recorded. The ion continues along its trajectory to a SSD, where its energy is recorded and secondary electrons are emitted from the surface of the SSD (actually, from a foil covering the surface). Those secondaries are accelerated by a grid into the same equipotential region, pass through the curved grid where they are reflected by the electric field between the second and third shells, and return through the equipotential region, through the bottom grid and into the same MCP, where the stop time is recorded. The Start position uniquely identifies the entrance aperture, and so the incident angle of the trajectory, the energy, and TOF combine to identify the ion species. The entrance foils are selected to be thick enough to suppress UV sufficiently to eliminate it as a source of background (a technique employed in many previous instruments).

# DEVELOPMENT

## *Internal*

Preliminary development of this sensor began under internal funding at APL. In particular, the electrostatic design was transferred to a computer-aided design (CAD) model, and plastic mock-up parts for that design fabricated using a fast-prototyping process. After plating with metal, we then bake these parts, assemble them, and test the electrostatics against breakdown in vacuum.

The figures below document the state of the hardware.

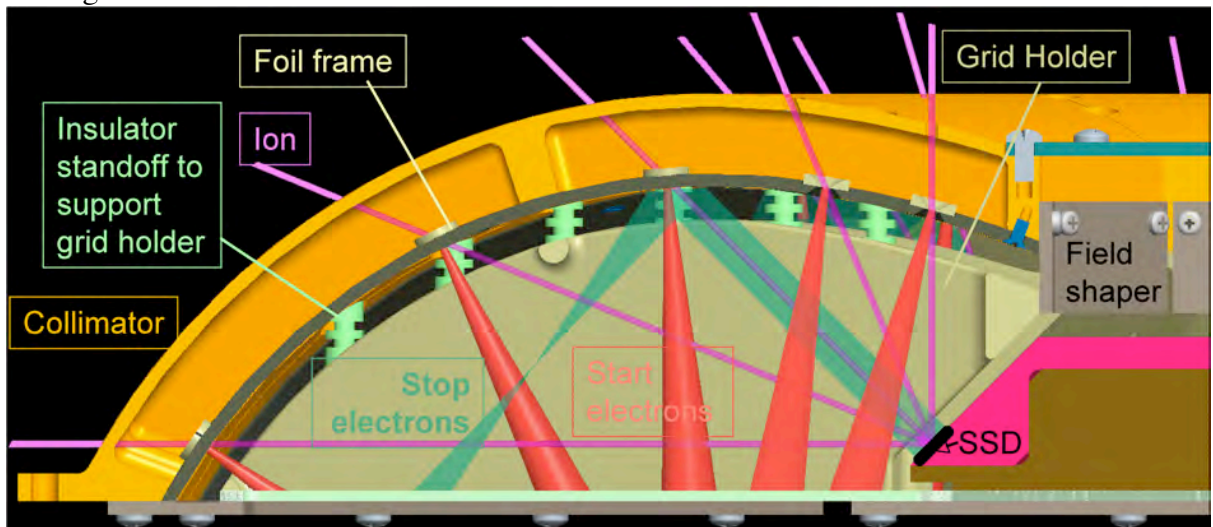

Figure 6. Cross-section of CAD model of electrostatics. Red cones illustrate Start-electron trajectories, green shading illustrates Stop-electron trajectories. Lavender lines illustrate incoming ion trajectories. SSD lies on the 45° pink surface to the right, where all the ion trajectories converge. Time dispersion for electron optics should not exceed ~100 pico-seconds system-wide.

After plating (except for the insulators), bake out, and assembly (including the grids and foils, which have not been included in the pre-plating assembly) we placed alpha-particle sources over the entrance foils (one at a time) and tested the secondary-electron-spot locations on a position-sensitive MCP located beneath the exit grid.

## *Plans at Start of NASA-Supported Funding*

Depending on the outcome of these prototype tests, we will proceed with further development of this sensor along appropriate tracks. If the tests show unexpected departures from the SIMION modeling, we will re-evaluate the model and the assembly, track down the reasons for the departure, correct them, and test again. One area that has not been rigorously modeled, but can be easily adjusted, is in the reflection of the Stop electrons. In the present design, that reflection takes place in a location that includes two of the entrance foils for the ions. While the lens elements are expected to remain sufficiently far from the electron reflection point that they introduce little perturbation to the electron trajectories, should the need arise we can adjust the angle of the SSD surface slightly, moving the reflection point farther out on the surface of the foil shell into a location distant from any of the entrance apertures.

Once these tests indicate the electrostatic design meets expectations for electron trajectories, we will test the electron time dispersion using a known, narrow-energy alpha source and lab electronics to measure particle time-of-flight for each of the ion-entrance foils in two separate octants. We will again make adjustments to the design as required by these results, with the goal of achieving <100 pico-second cumulative, electron-optics, time-dispersion.

When we are satisfied with the electrostatic design, we will concentrate on the design for the MCP timing and position detectors and their anodes, as well as on details the of the SSD mounting in the octagonal pyramid. The placement and function of these detectors is more critical in this sensor than in many past designs, which makes this phase of the development as important as the electrostatic design. The detectors must operate with low noise in the vicinity of high voltage electrostatic elements (grids, electrodes, etc.), and details of the mounting and of the mechanical design are important to achieving a working sensor.

The MCPs will be specially purchased shapes, to fit into each octant and receive all of the Start and Stop electrons from the foils in the sensor. The Start anodes comprise a series of discrete anodes connected in a daisy-chain by short delay lines (lumped delay lines, using LC components). A timing circuit receives signals from either end of the daisy-chained elements, and the relative delay observed between the ends identifies the particular discrete anode that was illuminated in an event. This reveals the particular (collimated) entrance foil transected by the ion, and so its trajectory. A second timing circuit measures the delay between the Start signal (corrected for the delay internal to the Start anode), and the Stop signal. This provides the particle TOF. When combined with the energy measured by the SSD, the particle mass can be determined.